# EFFECTIVE MULTI-STAGE TRAINING MODEL FOR EDGE COMPUTING DEVICES IN INTRUSION DETECTION

Trong Thua Huynh, Hoang Thanh Nguyen

Posts and Telecommunications Institute of Technology, Vietnam

## ABSTRACT

*Intrusion detection poses a significant challenge within expansive and persistently interconnected environments. As malicious code continues to advance and sophisticated attack methodologies proliferate, various advanced deep learning-based detection approaches have been proposed. Nevertheless, the complexity and accuracy of intrusion detection models still need further enhancement to render them more adaptable to diverse system categories, particularly within resource-constrained devices, such as those embedded in edge computing systems. This research introduces a three-stage training paradigm, augmented by an enhanced pruning methodology and model compression techniques. The objective is to elevate the system's effectiveness, concurrently maintaining a high level of accuracy for intrusion detection. Empirical assessments conducted on the UNSW-NB15 dataset evince that this solution notably reduces the model's dimensions, while upholding accuracy levels equivalent to similar proposals.*

## 1. INTRODUCTION

Over the past decade, numerous deep learning models have been proposed across various fields to replace previous traditional solutions, especially in network intrusion detection.Higher accuracy models tend to be more complex and larger in size. This leads to another significant challenge that needs to be addressed, which is how devices with limited resources and energy can detect network intrusions based on modern machine learning models.

As the Earth is getting warmer, alongside the effort to enhance problem-solving efficiency, we also need to ensure resource conservation, particularly energy resources, to safeguard a greener world. This study aims to achieve a balanced solution between intrusion detection effectiveness and resource consumption.

There are also many concentrated deep learning models within high-performance systems that provide results back to the end (customers). However, these models are still not widely adopted in a world where protecting customer data is important, and customers can build their own models with appropriate costs and goals. In return, their models need simplicity while ensuring high accuracy. Many deep learning models have generated astonishing prediction results with very high accuracy, but they require good enough resources such as high-speed, expensive GPUs. This can hardly be extensively applied to moderately configured hardware devices.

This research inherits the three-stage training model DSD that we published in [1]. In this study, we propose an enhanced three-stage training model, specifically tailored for edge computing devices. In the study [1], the DSD training model was divided into three stages, each utilizing a multi-layer deep learning model. We also experimented with the model using the UNSW-NB15 dataset, employing all three deep learning methods: RNN, LSTM [2], and GRU [3]. The results of that research showed that applying LSTM to the DSD training model yielded significantly superior prediction efficiency compared to RNN, GRU, and the original LSTM itself. Therefore, this current research focuses on improving the DSD model using LSTM, which we previously referred to as DSD-3hLSTM. We propose an enhanced pruning method combined with appropriate quantization to reduce the complexity of the model while maintaining high intrusion prediction accuracy.

The effectiveness of intrusion detection, especially when deployed on resource-constrained devices, cannot be overstated. These devices often grapple with constrained processing power, memory, and energy, making traditional intrusion detection methods challenging to implement. However, the integration of edge computing and the enhanced three-stage training model alleviates these limitations by distributing the computational workload intelligently. By processing data closer to the source, these devices can focus on initial data filtering and basic analysis, reducing the need for resource-intensive tasks. This streamlined approach conserves both computing power and energy, prolonging the operational life of these devices in today's interconnected digital landscape.

The structure of the paper is outlined as follows. Section 2 provides an overview of the related literature. In Section 3, we detail the hybrid model that we have developed. Moving forward, Sections 4 and 5 delve into the experimentation, results, and assessment of the proposed models. Ultimately, the paper concludes with final remarks in Section 6.

## 2. RELATED WORK

### 2.1. Long Short-Term Memory (LSTM)

The Long Short-Term Memory (LSTM) network, introduced as a variation of the recurrent neural network [2], stands as a pivotal machine learning technique tailored to address an array of sequential data challenges. Distinguished by its capacity to preserve and convey errors across temporal layers, LSTM substantially enhances the accuracy of output and imbues the recurrent neural network (RNN) with an extended capability for long-term memory tasks. The LSTM architecture, graphically depicted in Figure 1, encompasses four principal constituents: the input gate *(i)*, the forget gate *(f)*, the output gate *(o)*, and the memory cell *(c)*.

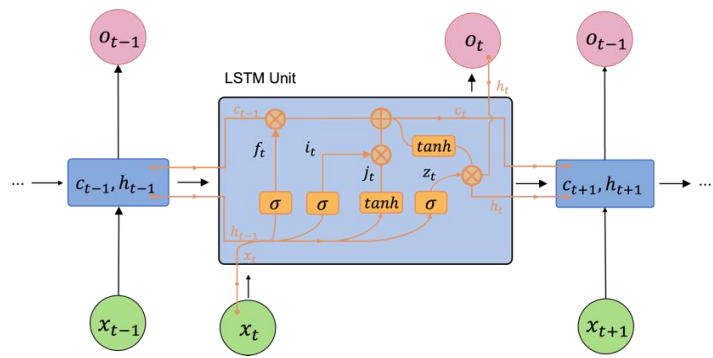

Figure 1. The core architecture of the LSTM block [2]

The LSTM block operates by determining the activation of its gates, which dictate the process of actions such as storage, reading, and writing, achieved through open or closed ports. Notably, each memory block corresponds to a discrete time step. The communication ports are established on a set of weights, with certain weights, such as those for input and hidden states, undergoing adjustments during the learning process. Equations ranging from (1) to (6) are employed to articulate the relationship between input and output at time *t* within each LSTM block.

$$f_t = \sigma(W_f[h_{t-1}, x_t] + b_f) \quad (1)$$
$$i_t = \sigma(W_i[h_{t-1}, x_t] + b_i) \quad (2)$$
$$j_t = \omega(W_j[h_{t-1}, x_t] + b_j) \quad (3)$$
$$c_t = f_t \times c_{t-1} + i_t \times j_t \quad (4)$$
$$z_t = \sigma(W_j[h_{t-1}, x_t] + b_j) \quad (5)$$
$$h_t = z_t \times \omega(c_t) \quad (6)$$

Where σ and ω are the activation functions *sigmoid* and *tanh*, respectively, $x_t$ is an input vector at time *t*, $h_t$ is the output vector at time *t*, *W* and *b* are the weight matrix and bias coefficient, respectively. $f_t$ is a forget function used to filter out unnecessary information, $i_t$ and $j_t$ are used to insert new information into memory cells, $z_t$ outputs relevant information.

## 2.2. Dense Sparse Dense (DSD)

Complex multi-layer neural architectures yield favorable outcomes by capturing highly nonlinear associations between input features and output. However, the drawback of these expansive models lies in their susceptibility to noise present in the training dataset, thereby giving rise to overfitting [4] and high variance [5]. Conversely, a reduction in model complexity can result in the omission of relevant relationships between features and output, leading to the predicament of underfitting [4] and high bias [6]. Achieving a harmonious balance between bias and variance is notably intricate and challenging.

Through the strategic application of pruning and network densification, the DSD training model introduces a transformative optimization approach. This, in turn, culminates in enhanced performance, yielding noteworthy outcomes.

## 2.3. Stochastic Gradient Descent (SGD)

Gradient Descent (GD) is an iterative optimization process aimed at searching for the optimal value of the objective function. It stands as one of the most widely used methods for adjusting model parameters to minimize the cost function within neural networks. Fundamentally, the GD algorithm is a prediction of θ with an initial point $\theta_0 = 0$. It then updates θ until an acceptable outcome $\theta = \theta - \eta \cdot \partial_\theta L(\theta)$, where $\partial_\theta L(\theta))$ denotes the derivative of the loss function *L* at $\theta$, and $\eta$ represents the stepsize (or learning rate).

Stochastic Gradient Descent (SGD) is a variant of the GD algorithm used to optimize machine learning models. It addresses the inefficiency in computation of traditional GD methods when handling large datasets in machine learning projects. In SGD, instead of using the entire dataset, it randomly selects a data point from the entire dataset at each iteration to compute the gradient and update the model parameters, thereby significantly reducing calculations.

Mathematically, the update rule of SGD after each epoch can be expressed as follows:

$$\theta = \theta - \eta \cdot \partial_\theta L(\theta; x^{(i)}; y^{(i)}) \quad (7)$$

where $\partial_\theta L(\theta; x^{(i)}; y^{(i)})$ is a loss function with only one data point pair (input, label) of $(x^{(i)}; y^{(i)})$.

To address the issue of SGD getting stuck in an undesirable local minimum, the Momentum method [7] is incorporated into SGD, resulting in a variant known as Stochastic Gradient Descent with Momentum (SGDM) [8]. The mathematical formulation for this approach is depicted as follows:

$$\theta = \theta - \eta \cdot \partial_\theta L(\theta; x^{(i)}; y^{(i)}) + \alpha \Delta_\theta \quad (8)$$

where α is an exponentialdecay factorbetween 0 and 1 that determines the relative contribution of the current gradient and earlier gradients to the weight change. $\Delta_\theta$ is the difference between two consecutive times of $\theta$.

SGDM is widely adopted for training neural networks, demonstrating notable experimental successes. It has been integrated as the default optimization algorithm in PyTorch [9] and TensorFlow [10], attesting to its practical significance and effectiveness.

## 2.4. Selective Weight Decay

*Regularization*, at its core, involves making slight adjustments to a model to mitigate overfitting while preserving its overall generalizability (the ability to describe a wide range of data, encompassing both training and test sets). In a more specific context, the aim is to guide the optimization problem's solution towards a nearby point. The direction of movement encourages the model to become less complex, even if it leads to a slight increase in the value of the loss function.

For Neural Network models, a commonly employed regularization technique is L2 regularization [11]. The essence of regularization involves augmenting the loss function with an additional term, denoted as $\mathcal{R}(\theta)$, as shown in the formula:

$$L_{reg}(\theta) = L(\theta) + \lambda \mathcal{R}(\theta) \quad (9)$$

This quantity influences the loss function. Specifically, when $\lambda$ is large, the impact of the added term on the loss function is substantial, whereas if $\lambda$ is small, the influence is minimal. However, $\lambda$ should not be excessively large, as an excessive value would overpower the added term. This causes the construction model to be wrong (underfitting). Since the weight parameters (w) primarily determine the predicted value delta, under L2 regularization, $\mathcal{R}$ is defined as:

$$\mathcal{R}(w) = \|w\|_2^2 \quad (10)$$

Consequently, the loss function can be expressed as:

$$L(w) = \frac{1}{2}\|y - Xw\|_2^2 + \lambda\|w\|_2^2 \quad (11)$$

The process of optimizing a model is synonymous with reducing the loss function, which in turn leads to a decrease in weights. Hence, L2 regularization is also referred to as 'weight decay,' as it contributes to the diminishment of weight values.

Selective Weight Decay (SWD) [12] is a continuous regularization process that induces sparsity in any kind of structure: at each training step, a specific penalty is applied to pruned parameters

International Journal of Computer Networks & Communications (IJCNC) Vol.16, No.1, January 2024

based on predefined criteria and a designated structure. The criterion for selection is the magnitude of weights or its derivatives according to the chosen structure. The optimization problem with penalties can be perceived as:

$$L(w) = \sum_{(x,y) \in D} \varepsilon(N(x,w), y) + \mu \|w\|_2 + a\mu \|w^*\|_2 \qquad (12)$$

where, $L$ is the objective function, the network is trained through error function $\varepsilon$ and penalized by a weight decay with a coefficient $\mu$ with $a$ being a coefficient determining the significance of SWD compared to the rest of the optimization problem, $w^*$ represents a subset of $w$ that is pruned at some step.

## 2.5. Quantization

Edge computing devices often come with constrained memory and computational power. Different optimization methods can be employed to adapt training models, enabling them to operate effectively within these limitations. Furthermore, specific optimization techniques are purposefully designed to leverage specialized hardware, thereby enhancing inference speed.
Several optimization techniques [13] can be utilized to reduce the model size. he small model offers many benefits such as reducing the memory capacity for training, thereby freeing up memory for other components, while potentially enhancing performance and stability. They also result in smaller download sizes, suitable for environments with limited bandwidth. Additionally, certain optimization approaches can decrease the required computational resources for inference, leading to lower latency. This, in turn, contributes to reduced power consumption.

However, optimization efforts can potentially lead to changes in the model's accuracy, a factor that must be carefully considered during the application's development process. The impact on accuracy varies depending on the specific model being optimized and is challenging to predict beforehand. In general, models optimized for size or latency may incur a marginal loss in accuracy. Depending on the application, a moderate reduction in accuracy might or might not significantly affect the end-user experience.

Quantization [14] is an optimization technique that facilitates the mapping of input values from a large set (often continuous) to output values in a smaller set (typically finite). It can reduce the precision of model parameters, but it concurrently diminishes latency and model size. According to this approach [15], real values $r$ can be derived from quantized values $q$ as follows:

$$r = S(q - Z) \qquad (13)$$

where, S and Z correspondingly represent the scale (utilized to shrink values with low precision back to floating-point values) and zero point (a value of low precision representing the quantized value, typically mapped to the real value 0).

This quantization method essentially combines the Quantize and Dequantize operations stacked upon each other. The scale is determined according to the formula:

$$scale = \frac{f_{max} - f_{min}}{q_{max} - q_{min}} \qquad (14)$$

where $f_{max}$ and $f_{min}$ represent the maximum and minimum values at the floating-point precision, while $q_{max}$ and $q_{min}$ denote the maximum and minimum values within the quantization range.

In this case, the zero point Z and the quantized value from the floating-point values are determined as follows:

$$Z = q_{min} - \frac{f_{min}}{scale} \tag{15}$$

$$q = \left\lfloor \frac{r}{S} + Z \right\rfloor \tag{16}$$

Thus, the real value is obtained by inserting the quantized value into equation (13), yielding:

$$r_{new} = S\left(\left\lfloor \frac{r}{S} + Z \right\rfloor - Z\right) \tag{17}$$

## 3. THE PROPOSAL

In this section, we present two parts. Part 1 (3.1) will outline the 3-stage DSD-3hLSTM model utilizing the LSTM deep learning method proposed in our earlier study [1]. Specifically, we will provide a detailed explanation of the pruning technique employed between hidden layers within the Sparse stage of the model. Part 2 (3.2) covers the improved pruning technique combined with quantization proposed for application to the DSD training model. The objective is to enhance training efficiency, reduce model complexity, while still maintaining a high level of accuracy in intrusion prediction.

### 3.1. Pruning in DSD-3hLSTM

Applying the findings of Song Han et al. in [16], in this training model, during the Sparse stage, connections with low weights are pruned, and subsequently, the network is trained after sparsification. A uniform sparsity is applied across all layers through a hyperparameter called sparsity, representing the percentage of weights to be pruned to zero. For each layer $W$ with $N$ parameters, the parameters are sorted, and the k-th largest parameter, denoted as $\lambda = S_k$, is chosen as the threshold, where $k = N * (1 - sparsity)$. Subsequently, all weights smaller than $\lambda$ are removed.

To eliminate small weights, a Taylor approximation polynomial is employed. The loss function and its Taylor polynomial are expressed in equations (18) and (19). To mitigate the loss increase when creating a hard threshold for weights, the first two components in formula (19) need to be minimized.

As parameters are reduced to zero, $\Delta W_i$ practically becomes $W_i - 0 = W_i$. At the local minimum level where $\partial L/\partial W_i \approx 0$ and $\frac{\partial^2 L}{\partial W_i^2} > 0$, only the second component is significant. Given that second-order gradient $\partial^2 L/\partial W_i^2$ is computationally expensive and $W_i$ is squared, $|W_i|$ is used as a pruning metric. A smaller $|W_i|$ implies a smaller increase in the loss function.

$$L = f(x, W_1, W_2, W_3 \dots) \tag{18}$$

$$\Delta L = \frac{\partial L}{\partial W_i} \Delta W_i + \frac{1}{2} \frac{\partial^2 L}{\partial W_i^2} \Delta W_i^2 + \cdots \tag{19}$$

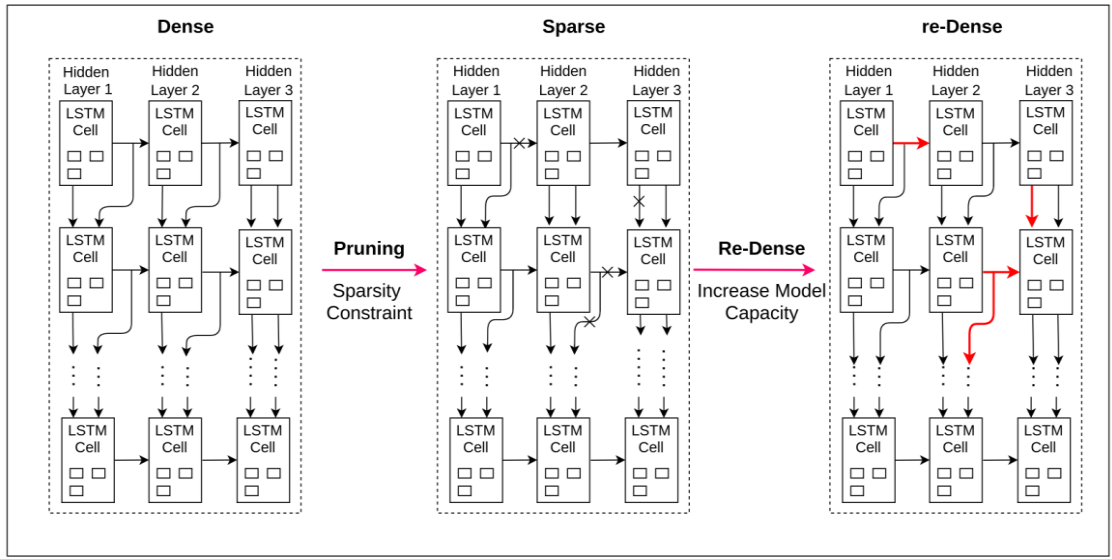

Figure 2. DSD-3hLSTM model

## 3.2. The Proposed Model

Building upon the DSD model proposed in our previous study [1], this research introduces improvements through a combined pruning technique and quantization method. The aim is to reduce model complexity, while maintaining a high level of accuracy in intrusion prediction.

**Pruning**

In this study, inspired by the momentum term in gradient descent learning algorithms proposed by Ning Qian in [7] and the Selective Weight Decay pruning technique introduced by Hugo Tessier et al. in [12], we present a three-stage training model named DSD-3hoLSTM. The purpose of this model is to perform continuous deep neural network pruning during the training process.

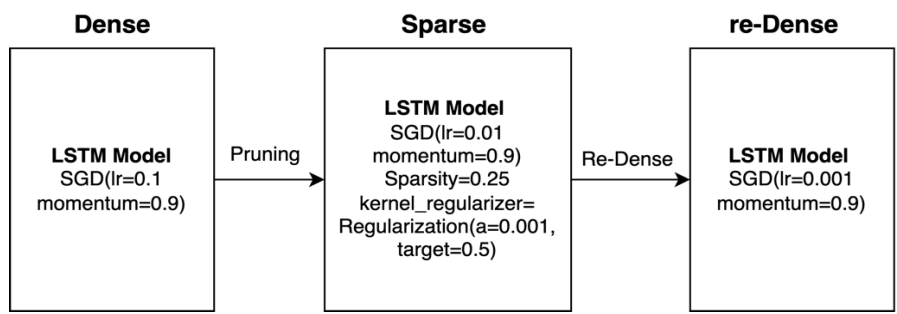

Figure 3. Model Optimization using SGDM

Initially, the model is optimized using the Stochastic Gradient Descent with Momentum (SGDM) method [8] to determine the optimal value of the objective function as described in equation (8) with specific parameter values detailed in Figure 3. In each stage, different learning rates are applied along with the same momentum. Specifically, in the Dense stage, SGDM is utilized to compute the optimal value of the loss function with a learning rate of 0.1. In the Sparse stage, the learning rate is set to 0.01, while in the final re-Dense stage, a learning rate of 0.001 is used. A consistent momentum of 0.9 is employed across all three stages.

Next, the model undergoes the pruning phase based on the Selective Weight Decay method [12], where we propose Algorithm 1 to enhance the training process in each stage. Specifically, in the First Phase and the Last Phase, we reuse the steps similar to the DSD algorithm described in our previous study [1]. However, in the Second Phase (Sparse), we introduce an improvement by adding a sub-mask layer to extract W* (a subset of W that has been pruned at some step). This subset contains weights greater than a target pruning threshold, denoted as *T*. We then compute the Total Weight Decay (TWD), which is the sum of all the regularization values after applying the mask and squaring the weights. Subsequently, we calculate *aTWD* by multiplying the total TWD with a coefficient *a* (determining the importance of TWD compared to the rest of the optimization problem). This step adjusts the level of regularization applied to the weights. Finally, the model updates its weights using the back propagate operation Err+WD+aTWD. Then, the *weight* values and *a* are updated to continue the iterative process until the target pruning threshold *T* is achieved.

Algorithm 1. DSD-3hoLSTM training procedure

```
Initialization: W^(0) with W^(0) N(0, Σ); a; T
Output: W^(t)
```
**The first Phase: Initialize Dense Phase**
    **while** not converged **do**
        $W^{(t)} = W^{(t-1)} - \eta^{(t)} \nabla f(W^{(t-1)}; x^{(t-1)})$
        $t = t + 1$
    **end**

**The second Phase: Sparse Phase**
    // initialize the mask by sorting and keeping the *k* weights at the top
    $S = sort(|W^{(t-1)}|);$
    $\lambda = S_k;$
    $Mask = 1(|W^{(t-1)}|) > \lambda;$
    **while** the network is not fully trained **do**
        $W^{(t)} = W^{(t-1)} - \eta^{(t)} \nabla f(W^{(t-1)}; x^{(t-1)})$
        $W^{(t)} = W^{(t)}.Mask$
        $Err \leftarrow \varepsilon(N(x^{(t-1)}, W^{(t)}), y^{(t-1)});$
        $WD \leftarrow \mu \|W^{(t)}\|_2;$
        $Mask' = 1(|W^{(t-1)}|) > a;$
        $W^{(t')} = W^{(t)}.Mask$
        determine $W^*$ according to $T$ and $W^{(t')}$;
        $TWD \leftarrow \mu \|W^*\|_2;$
        backpropagate $Err + WD + a.TWD$;
        update weights; increase $a$;
        $t = t + 1$
    **end**

**The last Phase: reDense Phase**
    **while** not converged **do**
        $W^{(t)} = W^{(t-1)} - \eta^{(t)} \nabla f(W^{(t-1)}; x^{(t-1)})$
        $t = t + 1$
    **end**

**Quantization**

After pruning, the model continues to undergo quantization according to the scheme depicted in Figure 4. From the DSD-3hLSTM model presented in our previous study [1], utilizing Stochastic Gradient Descent with Momentum, we obtain the DSD-3hoLSTM model (1.1). During the pruning step, the technique of weight sparsity is employed by retaining a certain number of important weights while setting the values of unimportant weights to 0. This reduction in model size not only decreases computational complexity but also enhances execution speed. By sparsifying the weights, we can generate smaller models suitable for deployment on resource-constrained infrastructures, such as mobile devices and embedded microcontrollers. Following this pruning step, we obtain a sparser model suitable for edge computing devices, denoted as DSD-3hoLSTM-pruned_model (2.1).

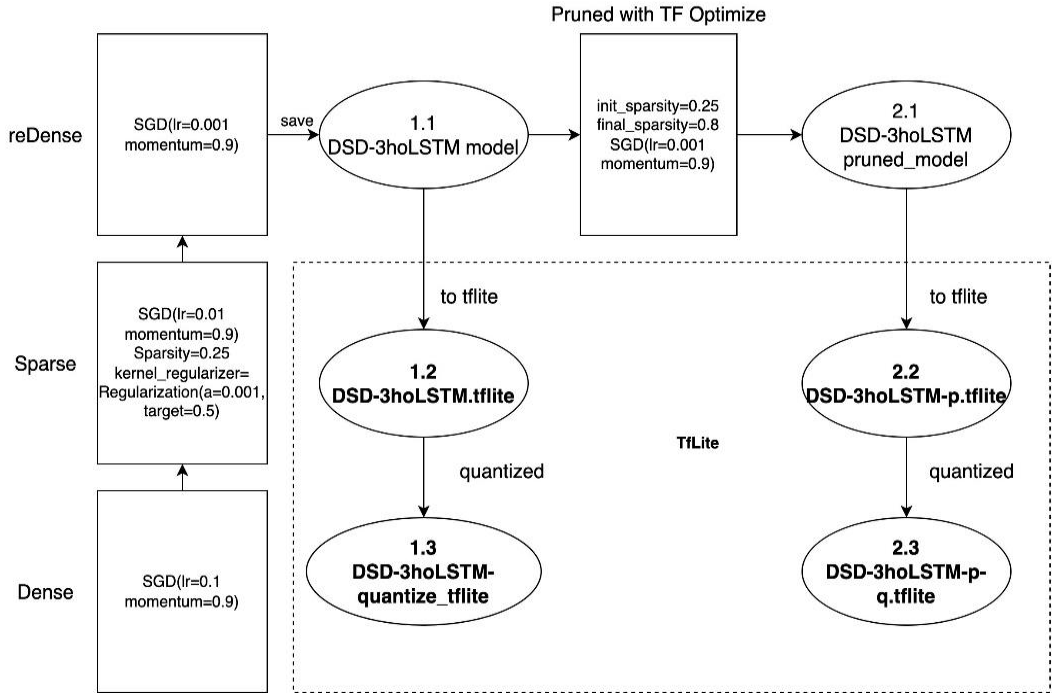

Figure 4. Quantization of the pruned model

In the quantization step, most of the values defined within the LSTM structure from equations (1) to (6) are easily quantizable, except for the value $c_t$ in equation (4). While the other equations utilize activation functions like tanh and sigmoid, $c_t$ does not. However, it is apparent that computations involving $c_t$ mainly consist of element-wise multiplications and additions, which may be much faster than matrix multiplications. For this reason, $c_t$ is represented in floating-point format, while the weights $W_f, W_i, W_j$ are quantized to the range of [-1, 1], and the input $x_t$ is quantized to the range of [0, 1].

In addition, due to the potential changes in weights and inputs over time or during the training process, we propose a quantization procedure based on the Dynamic Range Quantization method [17]. In this process, the range of numerical values is determined and converted into fixed-bit integer form, enhancing computational speed and reducing memory storage. Figure 5 outlines the steps of the dynamic range quantization execution, and the detailed implementation steps are presented in Algorithm 2. In the dynamic range quantization, weights are converted into precise 8-bit values. Consequently, from the original 32-bit values, this technique can reduce the model size by up to four times. The substantial reduction in model size is traded for minimal impact on latency and accuracy. In Figure 4, the model subjected to quantization is one that has been converted to the tflite format (1.2 and 2.2). Subsequently, we obtain the quantized model (1.3 and 2.3) after applying quantization techniques.

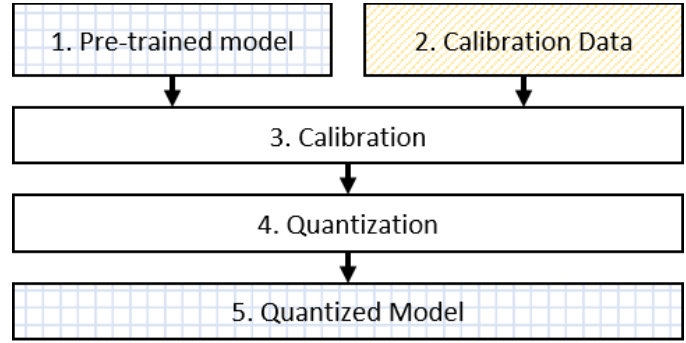

Figure 5. Dynamic Range Quantization Procedure

Algorithm 2. Dynamic Range Quantization

---

`Input`: Pre-trained model
  *Step 1.* **Calibration**
    *Step 1.1.* Find *min_value* and *max_value* based on the pre-trained model's weights and input data (optional).
    *Step 1.2.* Calculate the zero point *Z* and scale factor *S*, and quantize the values *q* from floating-point values using equations (14), (15), (16).
  *Step 2.* **Quantization**: Perform quantization for each weight in the model with fewer bits, based on the dynamic range determined in Step 1 using the formula (17).
`Output`: Quantized Model

---

## 4. EXPERIMENT

### 4.1. Dataset

In this study, to evaluate the effectiveness of the improved intrusion detection capability, we continue to use the UNSW-NB15 benchmark dataset [18], which was also utilized in the previous study [1]. This comprehensive dataset consists of a total of 2,540,044 records. The dataset described in the previous study [1] contains 49 features along with classified labels. These features encompass information such as total time (dur), protocol (proto), type of service (service), protocol state (state), number of packets from source to destination (spkts), number of packets from destination to source (dpkts), number of bytes from source to destination (sbytes), number of bytes from destination to source (dbytes), TTL value from source to destination (sttl), TTL value from destination to source (dttl), etc.

Indeed, the dataset includes additional features that provide more comprehensive information for intrusion detection. These features encompass:

- Source and destination jitter: Jitter refers to the variation in delay between packets in a network. It can provide insights into network congestion and quality.

- TCP connection establishment time: This feature indicates the time taken to establish a TCP connection between source and destination.
- Average stream packet size: This feature gives the average size of packets within a stream, helping to characterize the data flow.
- Link depth in the connection of HTTP request/response transaction: This feature provides information about the position of a request/response transaction within the connection.
- Number of connections with the same service and source/destination address: This feature counts the number of connections with the same service and source or destination address.
- Number of connections with the same source address and destination port: This counts connections with the same source address and destination port.

These additional features contribute to a more detailed representation of network traffic and interactions, enhancing the ability to detect intrusions and anomalous behavior effectively.

Based on the validated experimental results from the study [1], wechose the top 20% of the most important features from the dataset and apply custom features from UNSW-CF [19]. This decision was made to reduce training time while still ensuring effective intrusion detection. By selecting the most relevant features, the model can prioritize the information that contributes the most to detecting intrusions, leading to more efficient training and potentially faster inference during real-world usage.

### 4.2. Data Preprocessing

In the initial preprocessing step, we employed the scikit-learn LabelEncoder library [20] to transform nominal features into numerical representations before feeding them into the model.
In the UNSW-NB15 dataset, there are nine types of attacks (anomalies) and one additional class for normal traffic. The normal class contains 2,218,761 records, while the nine attack classes have a total of 321,283 records. These attack classes include *generics, exploits, fuzzers, DoS (Denial of Service), reconnaissance, analysis, backdoor, shellcode,* and *worms*. Detailed descriptions of these attack types are also provided in our previous paper [1].

Next, we employed the Min-Max normalization method to map the raw data values into the range of 0 to 1, facilitating more accurate operation of the objective function and enhancing the confidence level in predictions [21]. The Min-Max normalization equation is defined by formula:

$$x' = \frac{x - \min(x)}{\max(x) - \min(x)} \quad (20)$$

where $x$ represents the raw data value, and $x'$ denotes the normalized value.

### 4.3. Experimental Environment

The experiment was performed on a PC workstation, with a CPU configuration dual Intel Xeon E5-2683 2.1GHz, 64 GB memory, and GPU 8 Gigabytes Tesla P4. Experiments have been designed to study the performance of the DSD-3hLoSTM in binary classification (normal, anomaly). This model is trained on the comprehensive dataset UNSW-NB15, built on Python language and Keras library, Sklearn, runs on Tensorflow platform and Anaconda environment.The DSD-3hLoSTMinclude three phase Dense – Sparse – Redense. Each phase is designed as follow:

- The LSTM layers consist of 32 neuron units for all layer, all phases.

- The Dense layer uses the *sigmoid* activation function.
- Dropout layers have a dropout rate of 0.1 each layer hidden, all phases.
- SGD Optimization with learning rate is 0.1 for first phase, 0.01 for second phase, 0.001 for third phase. Momentum of 0.9 is applied for all phases.
- Dropout layers has a dropout rate of 0.1 each layer hidden.
- The initial_sparsity and final_sparsity parameters are experimented with values of 0.25 and 0.8, respectively.
- The coefficient a=0.001 and the threshold T=0.5.

## 5. RESULT EVALUATION

### 5.1. Evaluation Method

Selecting the appropriate evaluation method depends on the specific problem, our context, and the relative importance of minimizing false positives and false negatives. For instance, in an intrusion detection system, maximizing Detection Rate (recall) might be crucial to avoid missing potential intrusion cases, even if it results in more false alarms. In contrast, in a spam email filter, Precision might be more important to minimize false positives and prevent legitimate emails from being classified as spam. Intrusion detection systems often require a careful balance between the rate of true positive detections and the rate of false alarms.

We evaluate intrusion detection performance using 5 common metrics: Accuracy, Detection Rate, Precision, False Alarm Rate, and F1 score. Table 1 displays the confusion matrix, a table used to evaluate the performance of a classification model including True Positive (TP, i.e. Instances correctly classified as positive), True Negative (TN, i.e. Instances correctly classified as negative), False Positive (FP, i.e. Instances incorrectly classified as positive), and False Negative (FN, i.e. Instances incorrectly classified as negative). In our problem, TP and TN represent correctly classified attacked (anomaly) and normal states, respectively. FP indicates incorrect prediction of a normal record as an unrealistic attack, and FN indicates misclassification of an attack record as normal.

Table 1. Confusion Matrix

|  | **Prediction** – Normal (**0**) | **Prediction** – Anomaly (**1**) |
|---|---|---|
| Reality– **Normal (0)** | TN | FP |
| Reality– **Anomaly (1)** | FN | TP |

**Accuracy** – The formula calculating the accuracy is defined as:

$$Accuracy = \frac{TP + TN}{TP + TN + FP + FN} \quad (21)$$

**False Alarm Rate (FAR)** - This measure is calculated according to formula:

$$FAR = \frac{FP}{FP + TN} \quad (22)$$

**Precision** –The formula calculating Precision is defined as:

$$Precision = \frac{TP}{TP + FP} \quad (23)$$

24

**Detection Rate** (**DR** or **Recall**) –This criterion aims to evaluate the generalization of the found model and is determined by the formula:

$$DetectionRate = \frac{TP}{TP + FN} \qquad (24)$$

**F1-score**- is called a harmonic mean of the Precision and DR criteria and determined by the formula:

$$F1 - Score = \frac{2(Precision \times DR)}{Precision + DR} \qquad (25)$$

F1-score tends to take the value closest to whichever is the smaller between the Precision and DR values. Therefore, it is a more objective representation of the performance of a machine learning model.

## 5.2. Result and Evaluation

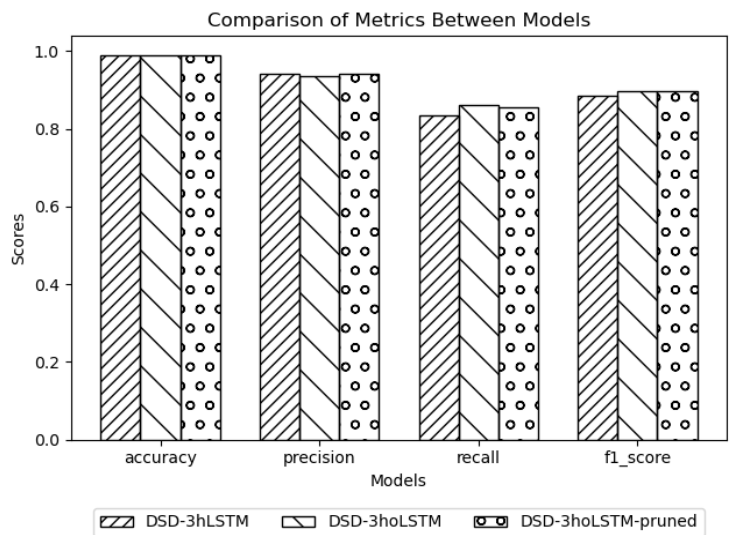

Figure 6. Visual Comparison of Model Similarity

Based on the results presented in Table 2, both proposed models in this study outperform the model from the study [1] in most metrics, except for the Precisionmetric. Additionally, the baseline model DSD-3hoLSTM performs better than the pruned model DSD-3hoLSTM-pruned in important metrics such as Accuracy, DR, and F1-score, although the differences are not significant. Observing Figure 6, we can clearly see a similarity in terms of accuracy among all three models, but the proposed models in this study exhibit better recall and f1-score values compared to the model from the study [1].

Table 2. Comparison between the proposed approach in this study and the approach in [1]

|  | FAR% | Acc% | Prec% | DR% | F1-score% |
|---|---|---|---|---|---|
| DSD-3hoLSTM | 0.2974 | **99.0526** | 93.6487 | **86.2710** | **89.8086** |
| DSD-3hoLSTM-pruned | 0.2690 | **99.0392** | 94.1668 | **85.4347** | **89.5885** |
| DSD-3hLSTM [1] | 0.2619 | 98.9540 | 94.1910 | 83.5346 | 88.5433 |

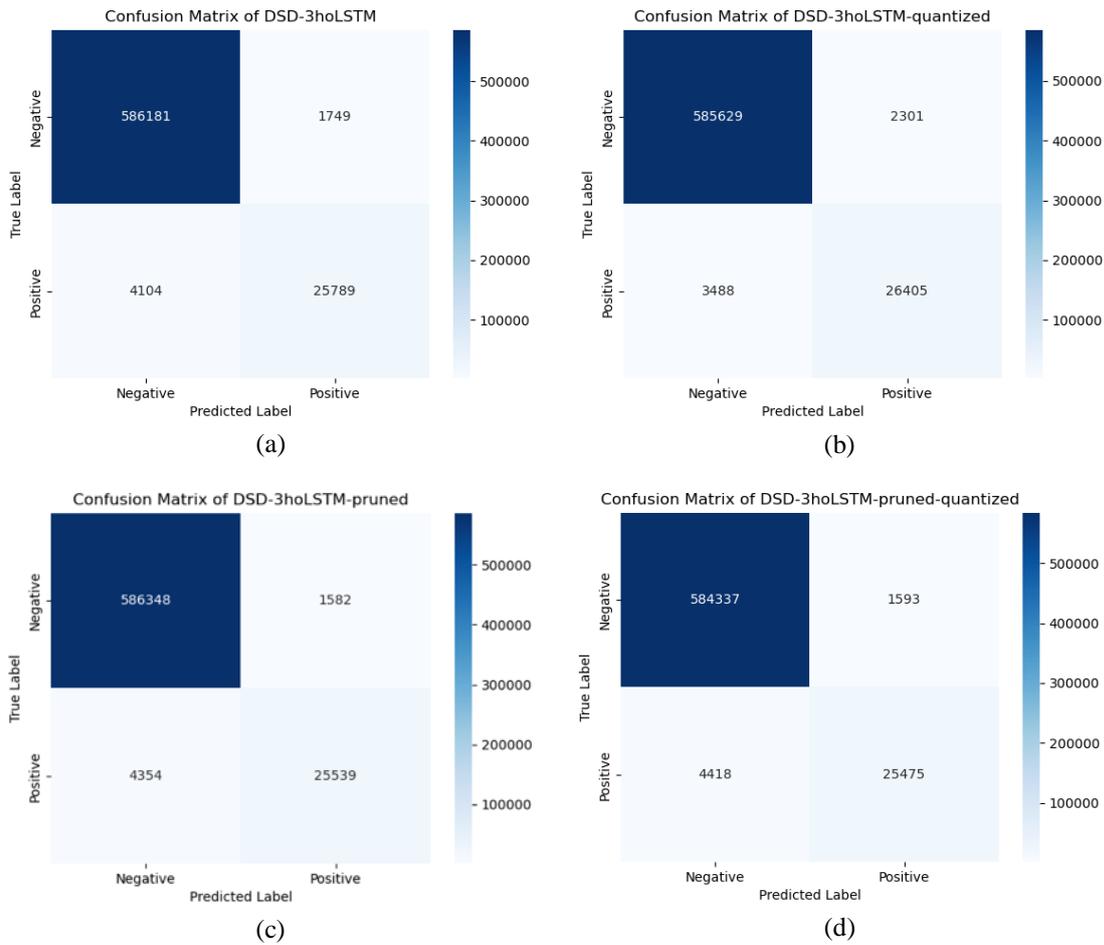

Figure 7. Confusion Matrix of the Models on the same Test Dataset

The Heatmap chart in Figure 7 displays confusion matrices based on the evaluation results of the four models: DSD-3hoLSTM, DSD-3hoLSTM-quantized, DSD-3hoLSTM-pruned, and DSD-3hoLSTM-pruned-quantized, all on the same test dataset. It is evident that the color patterns of correctly and incorrectly predicted regions are very similar among these models. The corresponding numerical values within these regions also reflect this high degree of similarity. Figure 7(d) exhibits the most noticeable differences compared to Figure 7(a), but the disparities in the numerical values are quite minimal. Additionally, Figures 7(b) and 7(c) show the closest resemblances to the other two plots. This observation suggests that pruning and quantization have a negligible impact on the models' ability to accurately predict correct or incorrect classifications.

Table 3. The combined pruning and quantization model vs individual models

|  | FAR% | Acc% | Prec% | DR% | F1-score% |
|---|---|---|---|---|---|
| 1.2 DSD-3hoLSTM | 0.2974 | 99.0526 | 93.6487 | 86.2710 | 89.8086 |
| 1.3 DSD-3hoLSTM-quantized | 0.3913 | **99.0630** | 91.9842 | **88.3317** | **90.1209** |
| 2.2 DSD-3hoLSTM-p | 0.2690 | 99.0392 | **94.1668** | 85.4347 | 89.5885 |
| 2.3 DSD-3hoLSTM-p-q | **0.2709** | 99.0270 | 94.1148 | 85.2206 | 89.4471 |

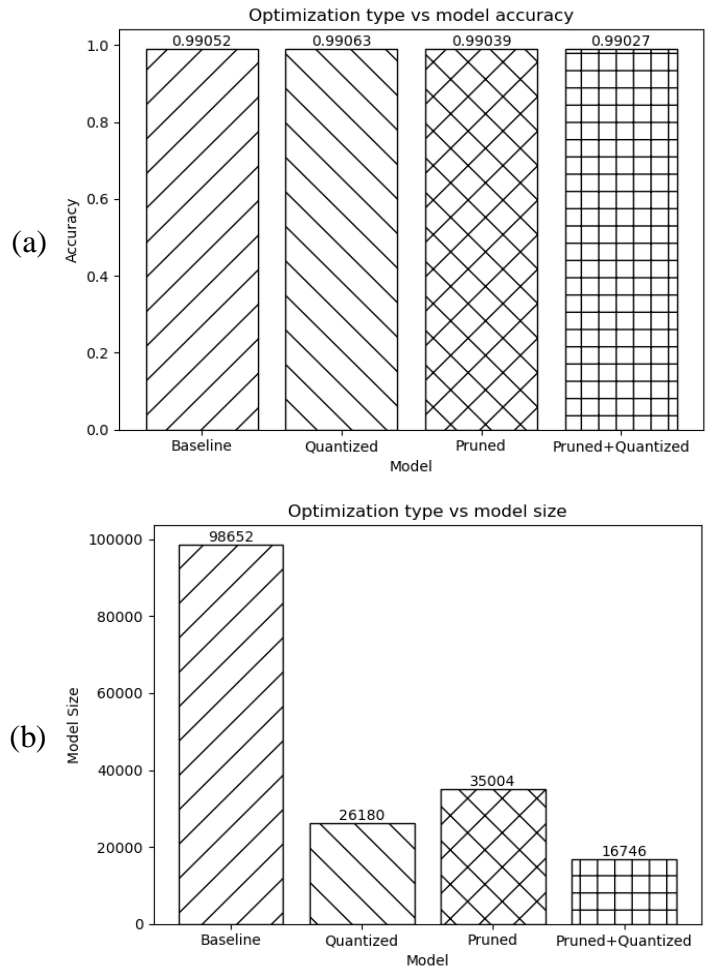

Figure 8. Visual Comparison of Model Similarity in Terms of Accuracy and Model Size

According to the results in table 3, the quantized, pruned, or combined models all outperform the baseline model in all metrics. Additionally, the quantized-only model performs better than the other models in terms of Accuracy, DR, and F1-score. The Precision metric is best achieved by the pruned-only model. The combined model, which includes both pruning and quantization, has the best FAR score, while other metrics such as Acc, Precision, DR, and F1-score show relatively similar results to both the quantized-only and pruned-only models. Therefore, to evaluate the model's effectiveness, we take more results on model size to compare in correlation with accuracy in order to recommend a suitable model when considering both these metrics.

In order to deploy the model in practical applications, we compared both the model size and accuracy when converting the DSD-3hoLSTM models (baseline), DSD-3hoLSTM-pruned, and quantized models into the tflite format using the Tensorflow platform. Figure 8 presents the results of evaluating the model's effectiveness based on two key metrics: prediction accuracy and model size. In Figure 8(a), the prediction accuracy of the models is nearly equivalent, with the highest being the quantized model (99.063%) and the lowest being the combined pruned-quantized model (99.027%). Thus, the difference in prediction accuracy between the highest and lowest is only 0.036%. Meanwhile, based on the data in Table 4 and the observation in Figure 8(b), we can easily notice a significant reduction in model size compared to the baseline model. The corresponding reduction ratios are approximately 2.81 times (65%), 3.76 times (73%), and 5.89 times (83%) for the pruned, quantized, and combined pruned-quantized models compared to

27

the baseline model. Clearly, with the approach proposed in this study, the DSD-3hoLSTM-pruned-quantized training model is well-suited for edge computing systems. The model size has been greatly reduced while maintaining a high level of prediction accuracy.

Table 4. Comparison of Model Effectiveness in Terms of Accuracy and Model Size

| Model in TFlite | Name | Accuracy (%) | Size (bytes) | +/-Size (times) |
|---|---|---|---|---|
| DSD-3hoLSTM.tflite | Baseline | 99.052 | 98,652 | |
| DSD-3hoLSTM-quantized.tflite | Quantized | 99.063 | 26,180 | -3.76 |
| DSD-3hoLSTM-pruned.tflite | Pruned | 99.039 | 35,004 | -2.81 |
| DSD-3hoLSTM-pruned-quantized.tflite | Pruned Quantized | 99.027 | 16,746 | -5.89 |

The True Positive Rate (TPR) is used to measure the proportion of positive cases correctly classified, while the False Positive Rate (FPR) is used to measure the proportion of positive cases incorrectly classified. The ROC curve represents the trade-off between TPR and FPR as the classification threshold changes from 0 to 1. Based on the ROC curve in Figure 9, the ROC curve of the proposed DSD-3hoLSTM model on both the validation and test datasets is better than that of the DSD-3hLSTM model [1]. Specifically, for the validation dataset, the ROC curve of DSD-3hoLSTM is 0.998753 compared to 0.963558 for DSD-3hLSTM. Similarly, on the test dataset, the ROC curve of DSD-3hoLSTM is 0.998823 compared to 0.998248 for DSD-3hLSTM. These results demonstrate the effectiveness of the DSD-3hoLSTM model compared to DSD-3hLSTM in early stopping to avoid overfitting and improving the model's prediction ability on new data.

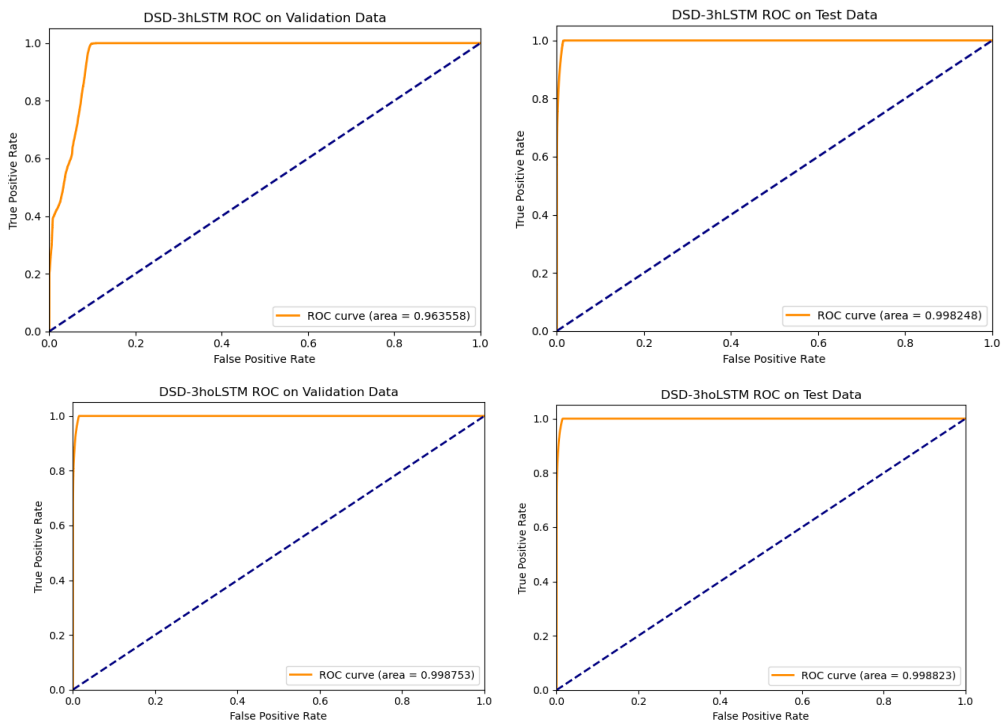

Figure 9. ROC Curve Comparison of the Models

## 6. CONCLUSIONS AND DISCUSSIONS

This paper presents a method to improve the efficiency of intrusion detection suitable for edge computing devices based on the combination of pruning and quantization techniques. This method not only ensures high accuracy in intrusion prediction, but also significantly reduces the size of the training model. To achieve that result, we have found a solution to the problem of local minimum and overfitting that exists in the DSD-3hLSTM model that we proposed previously based on Stochastic Gradient Descent with momentum and Selective Weight Decay methods. The significant contribution of this study is the improved pruning method, which eliminates many of the parameters that are not significant to the model. In addition, dynamic range quantization technique is also applied to the proposed model with suitable parameters to achieve the optimization of the training model. Through many experiments on the UNSW-NB15 dataset, the study has proven to be effective in intrusion detection with very high predictive accuracy while significantly reducing the model size. This shows the significance of the solution in creating suitable training models for edge computing devices.

Below are some further discussions on the trade-offs between model size and accuracy, especially in the context of pruning and quantization, with their implications for real-world implementations.

**Trade-offs between model size and accuracy**

Pruning techniques selectively remove weights, neurons, or entire layers from a neural network to reduce its size. This process often results in a smaller and sparser model. While pruning reduces model size significantly, it can lead to a reduction in accuracy, especially if the pruning criteria are not carefully chosen. Pruning might also introduce irregularities in the network structure, affecting its ability to generalize.

Quantization involves reducing the precision of model parameters, typically from 32-bit floating-point numbers to lower bit-width representations like 8-bit integers. Quantization can substantially decrease model size and memory footprint. However, lower precision can introduce quantization errors, potentially degrading model accuracy. The extent of accuracy loss depends on the specific quantization scheme and the type of data being processed.

**Implications for real-world deployment in many other areas**

Firstly, in edge computing scenarios, where devices have limited computational resources, smaller models resulting from pruning and quantization are advantageous. These models can run efficiently on edge devices, enabling real-time inference for applications like autonomous drones, industrial IoT, and wearable health monitors.Secondly, smaller models are beneficial for applications with limited bandwidth, such as remote sensing, where data must be transmitted over constrained networks. A smaller model size reduces the amount of data sent during inference, saving both time and data costs.

Finally, smaller models lead to faster inference times due to reduced computational requirements. This is crucial for real-time applications like autonomous vehicles, where split-second decisions are necessary for safety. Also, smaller models consume less power during inference, extending the battery life of mobile devices and reducing the environmental impact. This is essential for mobile applications, such as smartphone-based AI assistants.

CONFLICTS OF INTEREST

The authors declare no conflict of interest.

## AUTHORS

**Trong Thua Huynh** is currently the Head of Information Security Department, Faculty of Information Technology, Posts and Telecommunications Institute of Technology in Ho Chi Minh City, Vietnam. He received a Bachelor's degree in Information Technology from Ho Chi Minh City University of Natural Sciences, a Master degree in Computer Engineering at Kyung Hee University, Korea, and a Ph.D. degree in Computer Science at the Ho Chi Minh City University of Technology, Vietnam National University at Ho Chi Minh City. His key areas of research include Information Security in IoT, Blockchain, Cryptography, and Digital Forensics.

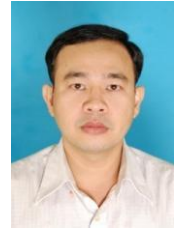

**Hoang Thanh Nguyen** is currently the Lecturer in Ho Chi Minh City, Vietnam. He received a Master's Degree in Information Systems from the Institute of Post and Telecommunications Technology. His research areas are Information Security, Machine Learning.

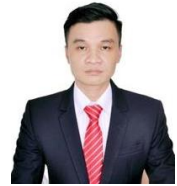